\documentclass[10pt,twocolumn,letterpaper]{article}
\usepackage[nocompress]{cite}
\usepackage{epsfig}
\usepackage{graphicx}
\usepackage{amsmath}
\usepackage{amssymb}
\usepackage{amsthm}
\usepackage{mathabx}
\usepackage{color} 

\usepackage[top=0.7in,bottom=0.7in,left=0.6in,right=0.7in,columnsep=0.35in]{geometry}
\date{\vspace{-0.3in}}
\newtheorem{theorem}{Theorem}
\newtheorem{lemma}[theorem]{Lemma}
\newcommand{\minus}[1]{{-#1}}
\DeclareMathOperator*{\argmin}{arg\,min}
\DeclareMathOperator*{\argmax}{arg\,max}
\newcommand{\E}{\mathbb{E}}
\renewcommand{\P}{\mathbb{P}}
\newcommand{\R}{\mathbb{R}}
\newcommand{\zero}{\mathbf{0}}

\renewcommand{\t}[1]{{#1}^{\rm T}}
\newcommand{\iverson}[1]{1[{#1}]}
\renewcommand{\exp}[1]{e^{#1}}

\newcommand{\si}[1]{^{(#1)}}
\newcommand{\greekbf}[1]{\text{\boldmath $#1$}}
\newcommand{\x}{\mathbf{x}}
\newcommand{\W}{\mathbf{W}}
\renewcommand{\O}{\mathcal{O}}
\newcommand{\NE}{{\mathcal{NE}}}
\newcommand{\G}{{\mathcal{G}}}
\newcommand{\Gt}{{\G^*}}
\newcommand{\qt}{{q^*}}
\renewcommand{\L}{\mathcal{L}}
\newcommand{\Gh}{\widehat{\G}}
\newcommand{\qh}{\widehat{q}}
\newcommand{\Gb}{\overline{\G}}
\newcommand{\qb}{\overline{q}}
\newcommand{\N}{\mathcal{N}}
\newcommand{\A}{\mathcal{A}}
\newcommand{\pih}{{\widehat{\pi}}}
\newcommand{\KL}{\mathbb{KL}}
\newcommand{\MI}{\mathbb{I}}
\newcommand{\VC}{\mathbb{VC}}
\newcommand{\PD}{\mathcal{P}}
\newcommand{\hns}{\hspace{-0.025in}}
\newcommand{\HS}{\mathcal{H}}
\newcommand{\QS}{\mathcal{Q}}
\newcommand{\DD}{\mathcal{D}}
\newcommand{\y}{\mathbf{y}}
\newcommand{\NEh}{\widehat{\NE}}
\newcommand{\NEt}{{\NE^*}}
\newcommand{\Gm}{{\G^-}}
\newcommand{\km}{{k^-}}
\newcommand{\NEm}{{\NE^-}}
\newcommand{\eps}{\varepsilon}
\newcommand{\BTheta}{\mathbf{\Theta}}
\newcommand{\vtheta}{\greekbf{\theta}}
\newcommand{\vphi}{\greekbf{\phi}}
\newcommand{\theoremspace}{\vspace{0.05in}}
\newcommand{\citep}[1]{\cite{#1}}

\newcommand{\citet}[1]{\cite{#1}}
\newcommand{\citealt}[1]{\cite{#1}}

\title{\textbf{On the Sample Complexity of Learning Graphical Games}}

\author{Jean Honorio \\
Computer Science, Purdue University \\
West Lafayette, IN 47907, USA \\
\texttt{jhonorio@purdue.edu}}

\begin{document}

\maketitle

\begin{abstract}

We analyze the sample complexity of learning graphical games from purely behavioral data.
We assume that we can only observe the players' joint actions and not their payoffs.
We analyze the sufficient and necessary number of samples for the correct recovery of the set of pure-strategy Nash equilibria (PSNE) of the true game.
Our analysis focuses on directed graphs with $n$ nodes and at most $k$ parents per node.
Sparse graphs correspond to ${k \in \O(1)}$ with respect to $n$, while dense graphs correspond to ${k \in \O(n)}$.
By using VC dimension arguments, we show that if the number of samples is greater than ${\O(k n \log^2{n})}$ for sparse graphs or ${\O(n^2 \log{n})}$ for dense graphs, then maximum likelihood estimation correctly recovers the PSNE with high probability.
By using information-theoretic arguments, we show that if the number of samples is less than ${\Omega(k n \log^2{n})}$ for sparse graphs or ${\Omega(n^2 \log{n})}$ for dense graphs, then any conceivable method fails to recover the PSNE with arbitrary probability.

\end{abstract}

\section{Introduction}

Non-cooperative game theory has been considered as the appropriate mathematical framework in which to formally study \emph{strategic} behavior in multi-agent scenarios.
The core solution concept of \emph{Nash equilibrium (NE)}~\citep{Nash51} serves a descriptive role of the stable outcome of the overall behavior of self-interested agents (e.g., people, companies, governments, groups or autonomous systems) interacting strategically with each other in distributed settings.

\vspace{-0.1in}
\paragraph{Algorithmic Game Theory and Applications.}

There has been considerable progress on \emph{computing} classical equilibrium solution concepts such as NE and \emph{correlated equilibria}~\citep{Aumann74} in graphical games (see, e.g., \citealt{Blum06,Kearns01,Jiang11,Kakade03,Ortiz02,Papadimitriou08,Vickrey02} and the references therein) as well as on computing the \emph{price of anarchy} in graphical games (see, e.g., \citealt{Benzwi11}).
Indeed, graphical games played a prominent role in establishing the computational complexity of computing NE in general normal-form games (see, e.g., \citealt{Daskalakis09} and the references therein).

In \emph{political science} for instance, the work of~\citep{Irfan14} identified the most influential senators in the U.S. congress (i.e., a small set of senators whose collectively behavior forces every other senator to a unique choice of vote).
The most influential senators were intriguingly similar to the gang-of-six senators formed during the national debt ceiling negotiation in 2011.
Additionally, it was observed in~\citep{Honorio15} that the influence from Obama to Republicans increased in the last sessions before candidacy, while McCain's influence to Republicans decreased.

\vspace{-0.1in}
\paragraph{Learning Graphical Games.}

The problems in \emph{algorithmic game theory} described above (i.e., computing Nash equilibria, computing the price of anarchy, or finding the most influential agents) require a known graphical game which is unobserved in the real world.
To overcome this issue, learning \emph{binary-action} graphical games from behavioral data was proposed in~\citep{Honorio15}, by using maximum likelihood estimation (MLE).
We also note that \citep{Honorio15,Irfan14} have shown the successful use of graphical games in real-world settings, such as the analysis of the U.S. congressional voting records as well as the U.S. supreme court.
More recently, the work of~\citep{Ghoshal16} provides a statistically and computationally efficient method for learning \emph{binary-action sparse} games.

\vspace{-0.1in}
\paragraph{Contributions.}

In this paper, we study the \emph{statistical} aspects of the problem of learning graphical games with \emph{general discrete} actions from strictly behavioral data.
As in~\citep{Ghoshal16,Honorio15}, we assume that we can only observe the players' joint actions and not their payoffs.
The class of models considered here are \emph{polymatrix graphical games}~\citep{Janovskaja68,Kearns01}.
We study the sufficient and necessary number of samples for the correct recovery of the pure-strategy Nash equilibria (PSNE) set of the true game, for directed graphs with $n$ nodes and at most $k$ parents per node.
Theorem~\ref{thm:suffpsne} shows that the sufficient number of samples for MLE is ${\O(k n \log^2{n})}$ for sparse graphs, and ${\O(n^2 \log{n})}$ for dense graphs.
Theorem~\ref{thm:necepsne} shows that the necessary number of samples for any conceivable method is ${\Omega(k n \log^2{n})}$ for sparse graphs, and ${\Omega(n^2 \log{n})}$ for dense graphs.
Thus, MLE is statistically optimal.

\vspace{-0.1in}
\paragraph{Discussion.}

While \emph{sparsity}-promoting methods were used in prior work \citep{Ghoshal16,Honorio15} for \emph{binary} actions, the benefit of sparsity for learning games with \emph{general discrete} actions has not been theoretically analyzed before.
In this paper, we focus on the \emph{statistical} analysis of exact MLE.\footnote{We leave the analysis of computationally efficient methods for future work.
To put this in context, note that theoretical analysis for learning Bayesian networks has focused exclusively on exact MLE~\citep{Brenner13}.}
Prior work on MLE estimation~\citep{Honorio15} has not focused on the correct PSNE recovery, but on generalization bounds.
More formally, Corollary~15 in~\citep{Honorio15} shows that for dense graphs with $n$ nodes and \emph{binary} actions, ${\O(n^3)}$ samples are sufficient for the empirical MLE minimizer to be close to the best achievable expected log-likelihood.
As a byproduct of our PSNE recovery analysis, Lemma~\ref{lem:gen} shows that for dense graphs and \emph{general discrete} actions, only ${\O(n^2 \log{n})}$ samples are sufficient for obtaining a good expected log-likelihood.
Regarding PSNE recovery, the results of~\citep{Ghoshal16} provide a ${\O(k^4 \log{n})}$ sample complexity for learning \emph{binary-action sparse} games.
The above results pertain to a specific class of payoff functions with a particular \emph{parametric} representation, that allows for a logistic regression approach.
The results in~\citep{Ghoshal16} also assume strict positivity of the payoffs in the PSNE set.
Thus, it is unclear how these results can be extended to \emph{general discrete} actions.

\section{Graphical Games}

In classical game-theory (see, e.g.~\citealt{Fudenberg91} for a textbook introduction), a \emph{normal-form game} is defined by a set of $n$ \emph{players} $V = \{1,\dots,n\}$, and for each player $i$, a set of \emph{actions}, or \emph{pure-strategies} $A_i$, and a payoff function $u_i : \A \to \R$ where $\A$ is the Cartesian product
\begin{align*}
\textstyle{ \A \equiv \bigtimes_{j \in V} A_j } \; .
\end{align*}
The payoff functions $u_i$ map the joint actions of all the players to a real number.
In non-cooperative game theory we assume players are greedy, rational and act independently, by which we mean that each player $i$ always want to maximize their own utility, subject to the actions selected by others, irrespective of how the optimal action chosen help or hurt others.

A core solution concept in non-cooperative game theory is that of an \emph{Nash equilibrium}.
A joint action $\x \in \A$ is a \emph{pure-strategy Nash equilibrium (PSNE)} of a non-cooperative game if, for each player $i$, $x_i \in \argmax_{a \in A_i} u_i(a,\x_\minus{i})$.
That is, $\x$ constitutes a \emph{mutual best-response}, no player $i$ has any incentive to unilaterally deviate from the prescribed action $x_i$, given the joint action of the other players ${\x_\minus{i} \in \bigtimes_{j \in V-\{i\}} A_j}$ in the equilibrium.
For normal-form games, we denote a game by ${\G = \{u_i : \A \to \R\}_{i \in V}}$, and the set of all \emph{pure-strategy Nash equilibria} of $\G$ by
\begin{align} \label{eq:psne}
\NE(\G) \equiv \{ \x \mid (\forall i \in V, a \in A_i){\rm\ }u_i(x_i,\x_\minus{i}) \geq u_i(a,\x_\minus{i}) \} \; .
\end{align}

A \emph{(directed) graphical game} is a game-theoretic graphical model~\citep{Kearns01}.
It provides a succinct representation of normal-form games.
In a graphical game, we have a (directed) graph $G = (V,E)$ in which each node in $V$ corresponds to a player in the game.
The interpretation of the edges/arcs $E$ of $G$ is that the payoff function of player $i$ is only a function of his own action and the actions of the set of parents/neighbors ${\N(i) \equiv \{ j \mid (i,j) \in E \}}$ in $G$ (i.e., the set of players corresponding to nodes that point to the node corresponding to player $i$ in the graph).
In this context, for each player $i$, we have a \emph{local} payoff function $u_i : A_i \times \bigtimes_{j \in \N(i)} A_j \to \R$.
A joint action $\x \in \A$ is a PSNE if, for each player $i$, ${x_i \in \argmax_{a \in A_i} u_i(a,\x_{\N(i)})}$.
For graphical games, we denote a game by ${\G = \{u_i : A_i \times \bigtimes_{j \in \N(i)} A_j \to \R\}_{i \in V}}$.

In this paper, we focus on \emph{polymatrix games}~\citep{Janovskaja68}.
Under this model, the local payoff functions ${u_i : A_i \times \bigtimes_{j \in \N(i)} A_j \to \R}$ have a succint representation as a sum of a unary potential function $u_{ii} : A_i \to \R$ and several pairwise potential functions $u_{ij} : A_i \times A_j \to \R$, that is
\begin{align} \label{eq:polymatrix}
u_i(x_i,\x_{\N(i)}) = u_{ii}(x_i) + \sum_{j \in \N(i)}{u_{ij}(x_i,x_j)} \; .
\end{align}
For polymatrix graphical games, we denote a game by ${\G = \{u_{ii} : A_i \to \R, u_{ij} : A_i \times A_j \to \R\}_{i,j \in V}}$.
We assume that $A_i$ is a countable finite set such that ${|A_i| \geq 2}$ for all players $i$.
Further, $|A_i| \in \O(1)$ with respect to $n$ and $k$.

The \emph{binary-action} models considered in~\citep{Ghoshal16,Honorio15,Irfan14} are a restricted subclass of the models that we consider here.
The results in~\citep{Ghoshal16,Honorio15,Irfan14} assume that $A_i = \{-1,+1\}$, $u_{ii}(x_i) = w_{ii} x_i$ and $u_{ij}(x_i,x_j) = w_{ij} x_i x_j$ for all $i,j$ and for a weight matrix $\W \in \R^{n \times n}$.

\vspace{-0.1in}
\paragraph{Equivalence Classes.}

Each PSNE set defines an \emph{equivalence class} of games for which players have the same joint behavior.
Thus, as argued further in Section 4 in~\citep{Honorio15} for \emph{binary-action} games, it is not possible to recover the structure and payoff functions of the true game from observed joint actions.
Instead, we can recover the PSNE set (or equivalence class) of the true game.
Here, we study the sufficient and necessary number of samples for the correct recovery of the PSNE set of the true game.

\vspace{-0.1in}
\paragraph{Main Assumptions.}

Our assumptions are minimal:
\begin{itemize}
\item We do not assume the availability of \emph{any} information regarding the structure or parameters of the \emph{true} graphical game.
  The problem is \emph{precisely} to infer that information.
\item We do not assume the availability of data related to the temporal \emph{dynamics}, i.e., each player's move.
  Instead, we assume that we only observe \emph{steady-state} joint actions, i.e., NE.
  Learning only from NEs, is arguably more challenging than learning from temporal dynamics.
\item To make learning even more challenging, we assume that data might be not entirely faithful to a graphical game.
  That is, we assume that a portion of the joint actions in the observed/training data is not an NE.
  This ``corruption'' can be modeled via a \emph{noise mechanism}.
\item Learning games is an \emph{unsupervised} task, i.e., we do not know which joint actions in the observed/training data are NE or not.
\item We assume that payoffs are unavailable in the observed/training data, which is a reasonable assumption in some real-world instances.
\end{itemize}

\section{Learning Graphical Games}

In this paper, we define $\HS$ to be the class of polymatrix graphical games with $n$ nodes and at most $k$ parents per node, as follows\footnote{$\NE(\G) \neq \emptyset$ and $\NE(\G) \neq \A$ ensure that $\PD_{\G,q}$ does not degenerate into a uniform distribution, see e.g., Definition~4 in~\citep{Honorio15}.}
\begin{align*}
\HS \equiv \left\{
\begin{array}{@{}l@{\hspace{0.04in}}l@{}}
\G \mid & \G = \{u_{ii} : A_i \to \R, u_{ij} : A_i \times A_j \to \R\}_{i,j \in V} \\
 & \wedge \, (\forall i \in V) {\rm\ } |\N(i)| \leq k \\
 & \wedge \, |\NE(\G)| \in \{1,\dots,|\A|-1\}
\end{array}
\right\} \; .
\end{align*}
Next, we introduce an extension of the generative model proposed in~\citep{Honorio15} originally for \emph{binary} actions.
Let $\G$ be a game, and let ${\QS_\G}$ be a set defined as follows\footnote{$q > |\NE(G)|/|A|$ ensures that $p_{\G,q}(\x_1)>p_{\G,q}(\x_2)$ for $\x_1 \in \NE(\G),\x_2 \notin \NE(\G)$, see e.g., Proposition~5 and Definition~7 in~\citep{Honorio15}.}
\begin{align*}
\textstyle{ \QS_\G \equiv \left( \frac{|\NE(\G)|}{|\A|} , 1-\frac{1}{2|\A|} \right] } \; .
\end{align*}
With some probability ${q \in \QS_\G}$, a joint action $\x$ is chosen uniformly at random from $\NE(\G)$;
otherwise, $\x$ is chosen uniformly at random from its complement set ${\A - \NE(\G)}$.
Hence, the generative model is a mixture model with mixture parameter
$q$ corresponding to the probability that a stable outcome (i.e., a
PSNE) of the game is observed, uniform over PSNE.
Formally, the probability mass function (PMF) over joint-behaviors $\x \in \A$ parameterized by $(\G,q)$ is
\begin{align} \label{eq:prob}
\textstyle{ p_{\G,q}(\x) \equiv q \, \frac{\iverson{\x \in \NE(\G)}}{|\NE(\G)|} + (1-q) \, \frac{\iverson{\x \notin \NE(\G)}}{|\A| - |\NE(\G)|} } \; ,
\end{align}
\noindent where we can think of $q$ as the ``signal'' level, and thus $1-q$ as the ``noise'' level in the data.
Additionally, ${\PD_{\G,q}}$ denotes the probability distribution defined by the PMF ${p_{\G,q}(\cdot)}$.

By using the PMF in eq.\eqref{eq:prob}, we can define a (scaled) negative log-likelihood function over joint-behaviors $\x \in \A$ for a game $\G$ and mixture parameter $q$ as follows
\begin{align} \label{eq:loglik}
\L_{\G,q}(\x) & = \textstyle{ - \frac{\log{p_{\G,q}(\x)}}{\log{(2|\A|^2)}} } \nonumber \\
 & = \textstyle{ - \frac{\iverson{\x \in \NE(\G)}}{\log{(2|\A|^2)}} \log{\frac{q}{|\NE(\G)|}} } \nonumber \\
 & \hspace{0.185in} \textstyle{ - \frac{\iverson{\x \notin \NE(\G)}}{\log{(2|\A|^2)}} \log{\frac{1-q}{|\A| - |\NE(\G)|}} } \; .
\end{align}
Note that since we scale the negative log-likelihood with a factor ${1/\log{(2|\A|^2)}}$ then ${\L_{\G,q}(\x) \in [0,1]}$ for all ${\G \in \HS}$, ${q \in \QS_\G}$ and ${\x \in \A}$.

Maximum likelihood estimation (MLE) allows to infer the game (and mixture parameter) from observed joint actions.
More formally, given a dataset $S$ of $m$ joint actions, the \emph{empirical} MLE minimizer is
\begin{align*}
(\Gh,\qh) = \argmin_{ \G \in \HS, q \in \QS_\G }{ \frac{1}{m} \sum_{\x \in S}{ \L_{\G,q}(\x) } } \; .
\end{align*}
Assume that a joint action $\x$ is drawn from an arbitrary data distribution $\DD$.
The \emph{expected} MLE minimizer is given by
\begin{align*}
(\Gb,\qb) = \argmin_{ \G \in \HS, q \in \QS_\G }{ \E_\DD[\L_{\G,q}(\x)] } \; .
\end{align*}
Note that if the data is generated by a \emph{true} game ${\Gt \in \HS}$ and mixture parameter ${\qt \in \QS_\Gt}$, then the expected MLE minimizer is the true game and mixture parameter.
That is, if ${\DD = \PD_{\Gt,\qt}}$ then ${\NE(\Gb)=\NE(\Gt)}$ and ${\qb=\qt}$.

\section{Sufficient Samples for PSNE Recovery}

In this section, we show that if the number of samples is greater than ${\O(k n \log^2{n})}$ for sparse graphs or ${\O(n^2 \log{n})}$ for dense graphs, then MLE correctly recovers the PSNE with high probability.

\vspace{-0.1in}
\paragraph{Number of PSNE Sets.}

First, we show that the number of PSNE sets induced by polymatrix graphical games is ${\O(\exp{k n \log^2{n}})}$ for sparse graphs, and ${\O(\exp{n^2 \log{n}})}$ for dense graphs.
These results will be useful later in obtaining a generalization bound as well as for analyzing the correct recovery of PSNE.
\theoremspace
\begin{lemma}[Number of PSNE sets] \label{lem:psne}
Let $\HS$ be the class of polymatrix graphical games with $n$ nodes and at most $k$ parents per node.
Let ${d(\HS)}$ be the number of PSNE sets that can be produced by games in $\HS$, i.e., ${d(\HS) = \left| \cup_{\G \in \HS} \{\NE(\G)\} \right|}$.
We have that ${d(\HS) \in \O(\exp{k n \log^2{n}})}$ for ${k \in \O(1)}$, and ${d(\HS) \in \O(\exp{n^2 \log{n}})}$ for ${k \in \O(n)}$.
\end{lemma}
\theoremspace
\begin{proof}
Let ${A_i = \{1,\dots,|A_i|\}}$ for all $i \in V$, w.l.o.g.
First, we introduce an equivalent representation of polymatrix graphical games.
To each unary potential function ${u_{ii} : A_i \to \R}$, we associate a vector ${\vtheta\si{i} \in \R^{|A_i|}}$ such that ${u_{ii}(x_i) = \sum_{b \in A_i}{\theta\si{i}_b \iverson{x_i=b}}}$.
To each pairwise potential function ${u_{ij} : A_i \times A_j \to \R}$, we associate a matrix ${\BTheta\si{i,j} \in \R^{|A_i| \times |A_j|}}$ such that ${u_{ij}(x_i,x_j) = \sum_{b \in A_i,c \in A_j}{\theta\si{i,j}_{bc} \iverson{x_i=b,x_j=c}}}$.
Note that the payoff functions $u_i$ are linear with respect to the vectors $\vtheta\si{i}$ and matrices $\BTheta\si{i,j}$ for all $i,j \in V$, that is
\begin{align*}
u_i(x_i,\x_\minus{i}) & = \sum_{b \in A_i}{\theta\si{i}_b \iverson{x_i=b}} \\
 & \hspace{0.15in} + \sum_{j \in V,b \in A_i,c \in A_j}{\hns\hns\hns\hns\hns \theta\si{i,j}_{bc} \iverson{x_i=b,x_j=c}} \; .
\end{align*}
In the above, we can define the parent/neighbor set ${\N(i) \equiv \{j \mid \BTheta\si{i,j} \neq \zero\}}$ and thus, summation across $j \in V$ is equivalent to summation across $j \in \N(i)$.

By eq.\eqref{eq:psne}, we have that a PSNE $\x$ fulfills ${u_i(x_i,\x_\minus{i}) - u_i(a,\x_\minus{i}) \geq 0}$ for all $i \in V$ and $a \in A_i$.
For polymatrix graphical games, for all players $i$ and $a \in A_i$, a PSNE $\x$ fulfills
\begin{align*}
\sum_{b \in A_i}{\theta\si{i}_b \iverson{x_i=b}} + \sum_{j \in V,b \in A_i,c \in A_j}{\hns\hns\hns\hns\hns \theta\si{i,j}_{bc} \iverson{x_i=b,x_j=c}} \\
- \theta\si{i}_a - \sum_{j \in V,c \in A_j}{\theta\si{i,j}_{ac} \iverson{x_j=c}} & \geq 0 \; .
\end{align*}
Thus, a PSNE is defined by $\sum_{i \in V}|A_i|$ \emph{linear} inequalities with respect to the vectors $\vtheta\si{i}$ and matrices $\BTheta\si{i,j}$.

For every player $i$ and $a \in A_i$, let ${D(i) \equiv (1+|A_i|)(1+\sum_{j \in V - \{i\}}{|A_j|})}$ and define the vectors ${\y\si{i,a} \in \{0,1\}^{D(i)}}$ and ${\vphi\si{i,a} \in \R^{D(i)}}$ as follows
\begin{align*}
\y\si{i,a} & \equiv (\{\iverson{x_i=b}\}_{b \in A_i}, \{\iverson{x_i=b,x_j=c}\}_{j \in V,b \in A_i,c \in A_j}, \\
 & \hspace{0.25in} -1, \{-\iverson{x_j=c}\}_{j \in V,c \in A_j}) \; , \\
\vphi\si{i,a} & \equiv (\{\theta\si{i}_b\}_{b \in A_i}, \{\theta\si{i,j}_{bc}\}_{j \in V,b \in A_i,c \in A_j}, \\
 & \hspace{0.25in} \theta\si{i}_a, \{\theta\si{i,j}_{ac}\}_{j \in V,c \in A_j}) \; .
\end{align*}

For polymatrix graphical games, a PSNE $\x$ fulfills ${\t{\vphi\si{i,a}}\y\si{i,a} \geq 0}$ for all $i \in V$ and $a \in A_i$.
Let ${D(i,k) \equiv (1+|A_i|)(1+\max_{\pi \subseteq V - \{i\}, |\pi| \leq k}{\sum_{j \in \pi}{|A_j|}})}$.
For every player $i$ and $a \in A_i$, define the function class ${\HS\si{i,a}}$ as follows
\begin{align*}
\HS\si{i,a} \equiv \left\{
\begin{array}{@{}l@{}}
f : \{0,1\}^{D(i)} \to \{0,1\} \, \mid \\
f(\y\si{i,a}) = \iverson{\t{\vphi\si{i,a}}\y\si{i,a} \geq 0} {\rm\ } \wedge \\
\vphi\si{i,a} \in \R^{D(i)} \, \wedge \, \sum_l{\iverson{\phi\si{i,a}_l \neq 0}} \leq D(i,k)
\end{array}
\right\} \; .
\end{align*}
Note that ${\HS\si{i,a}}$ is the class of linear classifiers in ${D(i)}$ dimensions, of weight vectors with at most ${D(i,k)}$ nonzero elements.
For ${k \in \O(1)}$, by Theorem~20 in~\citep{Neylon06} for the \emph{Vapnik-Chervonenkis (VC) dimension} of sparse linear classifiers, and since ${D(i,k) \in \O(k)}$ and ${D(i) \in \O(n)}$, the VC dimension of ${\HS\si{i,a}}$ is bounded as follows
\begin{align} \label{eq:vcsparse}
\VC(\HS\si{i,a}) & \leq 2(D(i,k)+1) \log{D(i)} \nonumber \\
 & \in \O(k \log {n}) \; .
\end{align}
For ${k = n-1}$, we have that ${D(i,n-1) = D(i) \in \O(n)}$, by the well-known VC dimension of linear classifiers, we have
\begin{align} \label{eq:vcdense}
\VC(\HS\si{i,a}) & \leq D(i)+1 \nonumber \\
 & \in \O(n) \; .
\end{align}

Recall that a PSNE is defined by ${D \equiv \sum_{i \in V}|A_i|}$ \emph{linear} inequalities.
Define the boolean function ${g : \{0,1\}^D \to \{0,1\}}$ as follows
\begin{align*}
\textstyle{ g(\{z_{ia}\}_{i \in V,a \in A_i}) \equiv \prod_{i \in V,a \in A_i}{z_{ia}} } \; .
\end{align*}
Note that if ${f\si{i,a} \in \HS\si{i,a}}$ for all $i$ and $a \in A_i$, then ${g(\{f\si{i,a}(\y\si{i,a})\}_{i \in V,a \in A_i})=1 {\rm\ } \Leftrightarrow {\rm\ } \x \in \NE(\G)}$.
Define the function class
\begin{align*}
g(\{\HS\si{i,a}\}_{i \in V, a \in A_i}) \equiv \left\{
\begin{array}{@{}l@{}}
g(\{f\si{i,a}(\y\si{i,a})\}_{i \in V,a \in A_i}) \, \mid \\
(\forall i \in V, a \in A_i) {\rm\ } f\si{i,a} \in \HS\si{i,a}
\end{array}
\right\} \; .
\end{align*}
By Lemma~2 in~\citep{Sontag98} then
\begin{align*}
\VC(g(\{\HS\si{i,a}\}_{i \in V, a \in A_i})) & \\
 & \hspace{-0.35in} \leq 2D (1+\log{D}) \max_{i \in V, a \in A_i}{\VC(\HS\si{i,a})} \; .
\end{align*}
Note that ${D \in \O(n)}$.
For ${k \in \O(1)}$, by eq.\eqref{eq:vcsparse}, we have ${\VC(g(\{\HS\si{i,a}\}_{i \in V, a \in A_i})) \in \O(k n \log^2{n})}$.
For ${k \in \O(n)}$, by eq.\eqref{eq:vcdense}, we have ${\VC(g(\{\HS\si{i,a}\}_{i \in V, a \in A_i})) \in \O(n^2 \log{n})}$.

Finally, note that our analysis of ${\VC(g(\HS_1,\dots,\HS_n))}$ provides a bound with respect to PSNE, while we are interested on PSNE sets.
Therefore, ${d(\HS) \leq \max_{i \in V}{|A_i|^{\VC(g(\HS_1,\dots,\HS_n))}} \in \O(\exp{\VC(g(\HS_1,\dots,\HS_n))})}$ and we prove our claim.
\qedhere
\end{proof}
\theoremspace

\vspace{-0.1in}
\paragraph{Generalization Bound.}

Next, we show that if the number of samples is greater than ${\O(k n \log^2{n})}$ for sparse graphs or ${\O(n^2 \log{n})}$ for dense graphs, then the empirical MLE minimizer is close to the best achievable expected log-likelihood.
\theoremspace
\begin{lemma}[Generalization bound] \label{lem:gen}
Fix ${\delta,\eps \in (0,1)}$.
Let $\HS$ be the class of polymatrix graphical games with $n$ nodes and at most $k$ parents per node.
Assume an arbitrary data distribution $\DD$.
Assume that $S$ is a dataset of $m$ joint actions (of the $n$ players), each independently drawn from $\DD$.
If ${m \in \O(\frac{1}{\eps^2}(k n \log^2{n} + \log{\frac{1}{\delta}}))}$ for ${k \in \O(1)}$ or ${m \in \O(\frac{1}{\eps^2}(n^2 \log{n} + \log{\frac{1}{\delta}}))}$ for ${k \in \O(n)}$, then
\begin{align*}
\P_S[ \E_\DD[\L_{\Gh,\qh}(\x) - \L_{\Gb,\qb}(\x)] \leq \eps ] \geq 1-\delta \; .
\end{align*}
\end{lemma}
\theoremspace
\begin{proof}
For clarity, let ${\L_S(\G,q) \equiv \frac{1}{m} \sum_{\x \in S}{ \L_{\G,q}(\x)}}$ and ${\L_\DD(\G,q) \equiv \E_\DD[\L_{\G,q}(\x)]}$.
By Lemma~11 in~\citep{Honorio15}, for any game $\G$ and for ${0 < q'' < q' < q < 1}$, if for any ${\eps > 0}$ we have
\begin{align*}
 & |\L_S(\G,q) - \L_\DD(\G,q)| \leq \eps \wedge |\L_S(\G,q'') - \L_\DD(\G,q'')| \leq \eps \\
 & \Rightarrow |\L_S(\G,q') - \L_S(\G,q')| \leq \eps \; .
\end{align*}
The above implies that for any game $\G$ and for any ${\eps>0}$, we have that
\begin{align*}
 & (\forall q \in \partial\QS_\G) {\rm\ } |\L_S(\G,q) - \L_\DD(\G,q)| \leq \eps \\
 & \Rightarrow (\forall q \in \QS_\G) {\rm\ } |\L_S(\G,q) - \L_\DD(\G,q)| \leq \eps \; ,
\end{align*}
\noindent where ${\partial\QS_\G}$ is the boundary of the set ${\QS_\G}$, i.e., ${\partial\QS_\G = \{\frac{|\NE(\G)|}{|\A|} , 1-\frac{1}{2|\A|}\}}$.
From the above, the union bound, the Hoeffding's inequality and Lemma~\ref{lem:psne}, we have that
\begin{align*}
\textstyle{ \P_S[(\forall \G \in \HS,q \in \QS_\G) {\rm\ } |\L_S(\G,q) - \L_\DD(\G,q)| \leq \frac{\eps}{2}] } \\
 & \hspace{-2.9in} = \textstyle{ 1 - \P_S[(\exists \G \in \HS,q \in \QS_\G) {\rm\ } |\L_S(\G,q) - \L_\DD(\G,q)| > \frac{\eps}{2}] } \\
 & \hspace{-2.9in} \geq \textstyle{ 1 - \P_S[(\exists \G \in \HS,q \in \partial\QS_\G) {\rm\ } |\L_S(\G,q) - \L_\DD(\G,q)| > \frac{\eps}{2}] } \\
 & \hspace{-2.9in} \geq \textstyle{ 1 - 2 \, d(\HS) \, \P_S[|\L_S(\G,q) - \L_\DD(\G,q)| > \frac{\eps}{2}] } \\
 & \hspace{-2.9in} \geq 1 - 4 \, d(\HS) \, \exp{-m \eps^2 / 2} \\
 & \hspace{-2.9in} \geq 1-\delta \; ,
\end{align*}
\noindent where ${d(\HS)}$ is the number of PSNE sets that can be produced by games in $\HS$, as defined in Lemma~\ref{lem:psne}.
The factor $2$ in ${2 \, d(\HS)}$ in the union bound comes from the fact that the set ${\partial\QS_\G}$ has exactly two elements.
Let ${T(n,k) \equiv k n \log^2{n}}$ if ${k \in \O(1)}$, and ${T(n,k) \equiv n^2 \log{n}}$ if ${k \in \O(n)}$.
By solving for $m$ in the last inequality, since ${d(\HS) \in \O(\exp{T(n,k)})}$, we get ${m \in \O(\frac{1}{\eps^2}(T(n,k) + \log{\frac{1}{\delta}}))}$.

We proved so far that with probability at least ${1-\delta}$, we have ${|\L_S(\G,q) - \L_\DD(\G,q)| \leq \frac{\eps}{2}}$ simultaneously for all ${\G \in \HS}$ and ${q \in \QS_\G}$.
Additionally, since ${(\Gh,\qh)}$ is the pair with minimum ${\L_S(\G,q)}$ from all ${\G \in \HS}$ and ${q \in \QS_\G}$, we have that
\begin{align*}
\E_\DD[\L_{\Gh,\qh}(\x) - \L_{\Gb,\qb}(\x)] & = \L_\DD(\Gh,\qh) - \L_\DD(\Gb,\qb) \\
 & \leq \textstyle{ \L_S(\Gh,\qh) + \frac{\eps}{2} - \L_S(\Gb,\qb) + \frac{\eps}{2} } \\
 & \leq \eps \; ,
\end{align*}
\noindent with probability at least ${1-\delta}$, which proves our claim.
\qedhere
\end{proof}
\theoremspace

\vspace{-0.1in}
\paragraph{Sufficient Samples for PSNE Recovery.}

Finally, we show that if the number of samples is greater than ${\O(k n \log^2{n})}$ for sparse graphs or ${\O(n^2 \log{n})}$ for dense graphs, then MLE correctly recovers the PSNE with high probability.
\theoremspace
\begin{theorem}[Sufficient samples for PSNE recovery] \label{thm:suffpsne}
Fix ${\delta,\eps \in (0,1)}$.
Let $\HS$ be the class of polymatrix graphical games with $n$ nodes and at most $k$ parents per node.
Assume that the data distribution ${\DD = \PD_{\Gt,\qt}}$ for some true game ${\Gt \in \HS}$ and mixture parameter ${\qt \in \QS_\Gt}$.
Assume that $S$ is a dataset of $m$ joint actions (of the $n$ players), each independently drawn from $\DD$.
If ${m \in \O(\frac{1}{\eps^2}(k n \log^2{n} + \log{\frac{1}{\delta}}))}$ for ${k \in \O(1)}$ or ${m \in \O(\frac{1}{\eps^2}(n^2 \log{n} + \log{\frac{1}{\delta}}))}$ for ${k \in \O(n)}$, then
\begin{align*}
\P_S[\NE(\Gt) \subseteq \NE(\Gh)] \geq 1-\delta \; .
\end{align*}
\noindent provided that ${|\NE(\Gt)| \geq 2}$ and ${\eps < \beta(|\NE(\Gt)|,\qt)}$ where\footnote{${\beta(r,q) \leq \frac{q}{2r}}$.
This maximum value is reached when ${|\A| \to \infty}$.}
\begin{align*}
\beta(r,q) & = \textstyle{ \frac{1}{\log{(2|\A|^2)}} \left(
\begin{array}{@{}l@{}}
q \log{\frac{q}{r}} + (1-q) \log{\frac{1-q}{|\A|-r}} \\
- \frac{r-1}{r} \, q \log{\frac{q}{r-1}} \\
- \left( \frac{q}{r} + 1-q \right) \log{\frac{1-q}{|\A|-r+1}}
\end{array}
\right) } \; .
\end{align*}
\end{theorem}
\theoremspace
\begin{proof}
Here, we follow a \emph{worst case} approach in which we analyze the identifiability of the PSNE set of $\Gt$ with respect to a game $\Gm$ that has one PSNE less than $\Gt$.
For our argument, showing the existence of such polymatrix graphical game $\Gm$ is not necessary.
In fact, a more general argument could be made with respect to a game that has  ${\km \geq 1}$ less PSNEs than $\Gt$.
The analysis for ${\km = 1}$ provides the sufficient conditions for the general case ${\km \geq 1}$.

For clarity, let ${c(n) \equiv \log{(2|\A|^2)}}$, ${\NEh \equiv \NE(\Gh)}$ and ${\NEt \equiv \NE(\Gt)}$.
Define the game $\Gm$ by its PSNE set ${\NEm \equiv \NE(\Gm)}$ as follows.
Define the set ${\NEm = \NEt - \{\x\}}$ for some ${\x \in \NEt}$.
It can be easily verified that
\begin{align*}
\begin{array}{@{}l@{\hspace{0.025in}}l@{\hspace{0.1in}}l@{\hspace{0.025in}}l@{}}
|\NEm| & = |\NEt|-1 \, , & \\
|\NEt \cap \NEm| & = |\NEt|-1 \, , & |\NEt \cup \NEm| & = |\NEt| \, , \\
|\NEt - \NEm| & = 1 \, , & |\NEm - \NEt| & = 0 \; .
\end{array}
\end{align*}
For any pair of games ${\G,\G' \in \HS}$, let ${\NE \equiv \NE(\G)}$ and ${\NE' \equiv \NE(\G')}$.
For any pair of games ${\G,\G' \in \HS}$, and mixture parameters ${q \in \QS_\G}$ and ${q' \in \QS_{\G'}}$, we have
\begin{align} \label{eq:halfkl}
\E_{\PD_{\G,q}}[\log{p_{\G',q'}(\x)}] & = \sum_{\x \in \A}{ p_{\G,q}(\x) \log{p_{\G',q'}(\x)} } \nonumber \\
 & \hspace{-0.55in} = \sum_{\x \in \NE \cap \NE'}{ p_{\G,q}(\x) \log{p_{\G',q'}(\x)} } \nonumber \\
 & \hspace{-0.4in} + \sum_{\x \in \NE - \NE'}{ p_{\G,q}(\x) \log{p_{\G',q'}(\x)} } \nonumber \\
 & \hspace{-0.4in} + \sum_{\x \in \NE' - \NE}{ p_{\G,q}(\x) \log{p_{\G',q'}(\x)} } \nonumber \\
 & \hspace{-0.4in} + \sum_{\x \notin \NE \cup \NE'}{ p_{\G,q}(\x) \log{p_{\G',q'}(\x)} } \nonumber \\
 & \hspace{-0.55in} = \textstyle{ \frac{|\NE \cap \NE'|}{|\NE|} \, q \log{\frac{q'}{|\NE'|}} } \nonumber \\
 & \hspace{-0.4in} + \textstyle{ \frac{|\NE - \NE'|}{|\NE|} \, q \log{\frac{1-q'}{|\A|-|\NE'|}} } \nonumber \\
 & \hspace{-0.4in} + \textstyle{ \frac{|\NE' - \NE|}{|\A|-|\NE|} \, (1-q) \log{\frac{q'}{|\NE'|}} } \nonumber \\
 & \hspace{-0.4in} + \textstyle{ \frac{|\A|-|\NE \cup \NE'|}{|\A|-|\NE|} \, (1-q) \log{\frac{1-q'}{|\A|-|\NE'|}} } \; .
\end{align}
Note that the pair ${(\Gm,\qt)}$ is well defined.
More formally, since $|\NEm| = |\NEt|-1$ then we have that ${\QS_\Gm = ( (|\NEt|-1)/|\A| , 1-1/(2|\A|) ]}$.
Thus, ${\qt \in \QS_\Gt \Rightarrow \qt \in \QS_\Gm}$.
From eq.\eqref{eq:halfkl}, we have
\begin{align*}
\KL(\PD_{\Gt,\qt} \| \PD_{\Gm,\qt}) & \\
 & \hspace{-0.3in} = \E_{\PD_{\Gt,\qt}}[\log{p_{\Gt,\qt}(\x)} - \log{p_{\Gm,\qt}(\x)}] \\
 & \hspace{-0.3in} = \textstyle{ \qt \log{\frac{\qt}{|\NEt|}} + (1-\qt) \log{\frac{1-\qt}{|\A|-|\NEt|}} } \\
 & \hspace{-0.15in} \textstyle{ - \frac{|\NEt|-1}{|\NEt|} \, \qt \log{\frac{\qt}{|\NEt|-1}} } \\
 & \hspace{-0.15in} \textstyle{ - \left( \frac{\qt}{|\NEt|} + 1-\qt \right) \log{\frac{1-\qt}{|\A|-|\NEt|+1}} } \; .
\end{align*}
By the assumption in the theorem and the above, we have that
\begin{align} \label{eq:epsGm}
c(n) \, \eps & < c(n) \, \beta(n,|\NEt|,\qt) \nonumber \\
 & = \KL(\PD_{\Gt,\qt} \| \PD_{\Gm,\qt}) \; .
\end{align}
Note that since ${\DD = \PD_{\Gt,\qt}}$ then ${\NE(\Gb)=\NE(\Gt)}$ and ${\qb=\qt}$.
By Lemma~\ref{lem:gen} and eq.\eqref{eq:loglik}, if ${m \in \O(\frac{1}{\eps^2}(k n \log^2{n} + \log{\frac{1}{\delta}}))}$ for ${k \in \O(1)}$ or ${m \in \O(\frac{1}{\eps^2}(n^2 \log{n} + \log{\frac{1}{\delta}}))}$ for ${k \in \O(n)}$, then
\begin{align*}
c(n) \, \eps & \geq c(n) \, \E_{\PD_{\Gt,\qt}}[\L_{\Gh,\qh}(\x) - \L_{\Gt,\qt}(\x)] \\
 & = \E_{\PD_{\Gt,\qt}}[\log{p_{\Gt,\qt}(\x)} - \log{p_{\Gh,\qh}(\x)}] \\
 & = \KL(\PD_{\Gt,\qt} \| \PD_{\Gh,\qh}) \; .
\end{align*}
Note that from the above and eq.\eqref{eq:epsGm}, we have that ${\KL(\PD_{\Gt,\qt} \| \PD_{\Gh,\qh}) < \KL(\PD_{\Gt,\qt} \| \PD_{\Gm,\qt})}$.
That is, the empirical MLE minimizer ${(\Gh,\qh)}$ is better than the pair ${(\Gm,\qt)}$.
Therefore, $\NEh$ includes all the PSNE in $\NEt$, i.e., ${\NEt \subseteq \NEh}$ and we prove our claim.
\qedhere
\end{proof}
\theoremspace

\vspace{-0.1in}
\paragraph{Remark.}

A similar argument as in Theorem \ref{thm:suffpsne} can be used to show that ${\NE(\Gh) \subseteq \NE(\Gt)}$, although the sufficient number of samples increases to ${\O(k n^3 \log^2{n})}$ for sparse graphs, and ${\O(n^4 \log{n})}$ for dense graphs.
(The function $\beta$ in such a case does not contain the ${1/\log{(2|\A|^2)} \in \O(1/n)}$ factor.)

\section{Necessary Samples for PSNE Recovery}

In this section, we show that if the number of samples is less than ${\Omega(k n \log^2{n})}$ for sparse graphs or ${\Omega(n^2 \log{n})}$ for dense graphs, then any conceivable method fails to recover the PSNE with probability at least $1/2$.
\theoremspace
\begin{theorem}[Necessary samples for PSNE recovery] \label{thm:necepsne}
Let $\HS$ be the class of polymatrix graphical games with $n$ nodes and at most $k$ parents per node.
Assume that the true game $\Gt$ is chosen uniformly at random (by nature) from a finite subset of $\HS$.
Assume that the true mixture parameter $\qt$ is known to the learner.
After choosing the true game $\Gt$, nature generates a dataset $S$ of $m$ joint actions (of the $n$ players), each independently drawn from ${\PD_{\Gt,\qt}}$.
Assume that a learner uses the dataset $S$ in order to choose a game $\Gh$.
If ${m \in \Omega(k n \log^2{n})}$ for ${k \in \O(1)}$ or ${m \in \Omega(n^2 \log{n})}$ for ${k \in \O(n)}$, then
\begin{align*}
\P_{\Gt,S}[\NE(\Gh) \neq \NE(\Gt)] \geq 1/2 \; ,
\end{align*}
\noindent for any conceivable learning mechanism for choosing $\Gh$.
\end{theorem}
\theoremspace
\begin{proof}
Let ${A_i = \{1,\dots,|A_i|\}}$ for all $i \in V$, w.l.o.g.
Let ${\Pi = \{ \pi \mid \pi \subseteq V \wedge |\pi| = k\}}$.
Let ${\pi \in \Pi}$ be the set of $k$ ``influential'' players.
Assume that nature picks $\pi$ uniformly at random from the ${\binom{n}{k}}$ elements in $\Pi$.
For a fixed $\pi$, we will construct a true game ${\G^\pi}$.
For clarity, we define ${\G^\pi \equiv \Gt}$ and ${q \equiv \qt}$.
The goal of the learner is to use the dataset $S$ in order to choose a set $\pih$ of $k$ players, and to output a game ${\G^\pih \equiv \Gh}$.

For a fixed $\pi$, we construct a game ${\G^\pi}$ with a single PSNE (i.e., ${|\NE(\G^\pi)|=1}$) as follows.
The $k$ ``influential'' players do not have any parent, i.e., $\N(i) = \emptyset$ for $i \in \pi$.
We force the ``influential'' players $i \in \pi$ to have a best response $1$, by setting their potential functions as follows.
\begin{align*}
(\forall i \in \pi) {\rm\ } u_{ii}(x_i) = \iverson{x_i = 1} \; .
\end{align*}
By eq.\eqref{eq:polymatrix}, the local payoff function for $i \in \pi$ becomes $u_i(x_i) = \iverson{x_i = 1}$.
The remaining ${n-k}$ ``influenced'' players have the $k$ ``influential'' players as parents, i.e., $\N(i) = \pi$ for $i \notin \pi$.
We force the ``influenced'' players to have a best response $2$, by setting their potential functions as follows
\begin{align*}
(\forall i \notin \pi) {\rm\ } u_{ii}(x_i) & = 0 \; , \\
(\forall i \notin \pi, j \in \pi) {\rm\ } u_{ij}(x_i,x_j) & = \iverson{x_i = 2, x_j = 1} \; .
\end{align*}
By eq.\eqref{eq:polymatrix}, the local payoff function for $i \notin \pi$ becomes $u_i(x_i,x_{\N(i)}) = \sum_{j \in \pi}{\iverson{x_i = 2, x_j = 1}}$.
The constructed game ${\G^\pi}$ has a single PSNE $\x^\pi$.
More specifically
\begin{align*}
(\forall i \in \pi) {\rm\ } x^\pi_i & = 1 \; , \\
(\forall i \notin \pi) {\rm\ } x^\pi_i & = 2 \; , \\
\NE(\G^\pi) & = \{ \x^\pi \} \; .
\end{align*}
Since we assume a known fixed mixture parameter $q$ and since ${|\NE(\G^\pi)|=1}$, the PMF defined in eq.\eqref{eq:prob} reduces to
\begin{align*}
p_\pi(\x) & \equiv p_{\G^\pi,q}(\x) \\
 & = \textstyle{ \iverson{\x = \x^\pi} \, q + \iverson{\x \neq \x^\pi} \, \frac{1-q}{|\A| - 1} } \; .
\end{align*}
Let ${\PD_\pi}$ denote the probability distribution defined by the PMF ${p_\pi(\cdot)}$.
Clearly, ${\pi \neq \pi' {\rm\ } \Leftrightarrow {\rm\ } \x^\pi \neq \x^{\pi'}}$.
Thus, for all ${\pi \neq \pi'}$ the Kullback-Leibler divergence is bounded as follows
\begin{align*}
\KL(\PD_\pi \| \PD_{\pi'}) & = \sum_{\x \in \A}{p_\pi(\x) \log{p_\pi(\x)} } - \sum_{\x \in \A}{p_\pi(\x) \log{p_{\pi'}(\x)} } \\
 & \hspace{-0.1in} = p_\pi(\x^\pi) \log{p_\pi(\x^\pi)} + \sum_{\x \neq \x^\pi}{ p_\pi(\x) \log{p_\pi(\x)} } \\
 & \hspace{0.05in} - p_\pi(\x^\pi) \log{p_{\pi'}(\x^\pi)} - p_\pi(\x^{\pi'}) \log{p_{\pi'}(\x^{\pi'})} \\
 & \hspace{0.05in} - \sum_{\x \notin \{\x^\pi,\x^{\pi'}\}}{p_\pi(\x) \log{p_{\pi'}(\x)} } \\
 & \hspace{-0.1in} = \textstyle{ q \log{q} + (|\A|-1) \frac{1-q}{|\A|-1} \log{\left(\frac{1-q}{|\A|-1}\right)} } \\
 & \hspace{0.05in} \textstyle{ - q \log{\left(\frac{1-q}{|\A|-1}\right)} - \frac{1-q}{|\A|-1} \log{q} } \\
 & \hspace{0.05in} \textstyle{ - (|\A|-2) \frac{1-q}{|\A|-1} \log{\left(\frac{1-q}{|\A|-1}\right)} } \\
 & \hspace{-0.1in} = \textstyle{ \frac{|\A| q - 1}{|\A|-1} \left(\log{q} - \log{\left(\frac{1-q}{|\A|-1}\right)} \right) } \; .
\end{align*}
Assume that the value of the mixture parameter (known to the learner) is ${q \equiv 2/|\A| \in \QS_{\G^\pi}}$.
Thus, for all ${\pi \neq \pi'}$ we have
\begin{align*}
\KL(\PD_\pi \| \PD_{\pi'}) & = \textstyle{ \frac{\log{(|\A|-1)} - \log{(|\A|/2-1)}}{|\A|-1} } \\
 & \in \O(1/(n \log{n}) \; .
\end{align*}
Conditioned on $\pi$, $S$ is a dataset of $m$ i.i.d. joint actions drawn from ${\PD_\pi}$.
That is, ${S \mid \pi \sim \PD_\pi^m}$.
The mutual information can be bounded by a pairwise KL-based bound~\citep{Yu97} as follows
\begin{align*}
\MI(\pi,S) & \leq \frac{1}{|\Pi|^2} \sum_{\pi \in \Pi}{\sum_{\pi' \in \Pi}{ \KL(\PD_\pi^m \| \PD_{\pi'}^m) }} \\
 & \leq \max_{\pi \neq \pi'}{ \KL(\PD_\pi^m \| \PD_{\pi'}^m) } \\
 & = m \, \max_{\pi \neq \pi'}{ \KL(\PD_\pi \| \PD_{\pi'}) } \\
 & \in \O(m/(n \log{n})) \; .
\end{align*}
Note that ${\pih = \pi {\rm\ } \Leftrightarrow {\rm\ } \NE(\G^\pih) = \NE(\G^\pi)}$.
Let ${T(n,k) \equiv k \log{n}}$ if ${k \in \O(1)}$, and ${T(n,k) \equiv n}$ if ${k = n/2}$.
Next, we show that ${\log{|\Pi|} \in \Omega(T(n,k))}$.
For ${k \in \O(1)}$, we have ${|\Pi| = \binom{n}{k} \geq (\frac{n}{k})^k}$ and thus ${\log{|\Pi|} \in \Omega(k \log{n}) = \Omega(T(n,k))}$.
For ${k = n/2}$, we have ${|\Pi| = \binom{n}{n/2} \geq (\frac{n}{n/2})^{n/2} = 2^{n/2}}$ and thus ${\log{|\Pi|} \in \Omega(n) = \Omega(T(n,k))}$.
By the Fano's inequality~\citep{Cover06} on the Markov chain ${\pi \to S \to \pih}$ we have
\begin{align*}
\P_{\Gt,S}[\NE(\Gh) \neq \NE(\Gt)] & = \P_{\pi,S}[\NE(\G^\pih) \neq \NE(\G^\pi)] \\
 & = \P_{\pi,S}[\pih \neq \pi] \\
 & \geq 1 - \frac{ \MI(\pi,S) + \log{2} }{ \log{|\Pi|} } \\
 & \geq 1 - \O\left( \frac{ m/(n \log{n}) }{ T(n,k) } \right) \\
 & = 1/2 \; .
\end{align*}
By solving the last equality, we prove our claim.
\qedhere
\end{proof}

\section{Concluding Remarks}

There are several ways of extending this research.
Other noise processes can be analyzed, such as a local noise model where the observations are drawn from the PSNE set, and subsequently, each action is independently corrupted by noise.
Other equilibria concepts can also be studied, such as mixed-strategy Nash equilibria, correlated equilibria and epsilon Nash equilibria.

\paragraph{Acknowledgements.}

We thank Xi Chen and Richard Cole for the helpful and valuable discussions.

\bibliography{references}

\end{document}